\documentstyle[epsfig,12pt]{aipproc}
\begin{document}

\title{Total Photonic and Hadronic Cross-sections}
\author{
 R.M. Godbole$^*$, A. Grau$^\dagger$, G. Pancheri$^{**}$}
\address{
$^*$Centre for Theoretical Studies, 
                    Indian Institute of Science, Bangalore 560 012, India \\
$^\dagger$ Departamento de F\'\i sica Te\'orica y del Cosmos, Universidad de
Granada, Spain\\ 
$^{**}$ INFN - Laboratori Nazionali di Frascati, 
Via E. Fermi 40, I00044 Frascati, Italy}

\begin{flushright}
IISc-CTS/26/00 \\
UG-FT-123/00 \\
LNF-00/027(P) \\
hep-ph/0101321 
\end{flushright}

\vskip 25pt
\begin{center}

{\large\bf Total Photonic and Hadronic Cross-sections \footnote{Talk 
presented  by G.P. at Photon-2000, Ambelside, U.K., Aug 26-31, 2000}}    
       \\
\vskip 25pt

{\bf                        Rohini M. Godbole } \\ 

{\footnotesize\rm 
                      Centre for Theoretical Studies, 
                     Indian Institute of Science, Bangalore 560 012, India. \\ 
                     E-mail: rohini@cts.iisc.ernet.in  } \\ 

\vskip 20pt

{\bf                                 A. Grau} \\

{\footnotesize\rm 
             Departamento de F\'\i sica Te\'orica y del Cosmos, \\
               Universidad de Granada, Spain\\ 
               E-mail: igrau@ugr.es } \\

\vskip 20 pt

{\bf                              G. Pancheri} \\

{\footnotesize\rm 
                    INFN - Laboratori Nazionali di Frascati, \\
                    Via E. Fermi 40, I00044 Frascati, Italy\\
                    E-mail: Giulia.Pancheri@lnf.infn.it }\\

\vskip 20pt

{\bf                             Abstract 
}

\end{center}

\begin{quotation}
\noindent
We discuss total cross-sections within the context of the QCD calculable
mini-jet model, highlighting its successes and failures. In particular
we show its description   of $\gamma \gamma \rightarrow hadrons$ and
compare it with OPAL and L3 data. We extrapolate this result to
$\gamma\ p$ total cross-sections and propose a phenomenological
ans\"atz for virtual photon cross-sections. We point out that the good
 agreement with data 
obtained with the Eikonal Minijet Model should not hide the 
many uncertainties buried in the impact parameter distribution.
A model obtained from Soft Gluon Summation is
briefly discussed and its application to hadronic cross-sections 
is shown.
\end{quotation}
 
\newpage
\maketitle

\begin{abstract}
We discuss total cross-sections within the context of the QCD calculable
mini-jet model, highlighting its successes and failures. In particular
we show its description   of $\gamma \gamma \rightarrow hadrons$ and
compare it with OPAL and L3 data. We extrapolate this result to
$\gamma\ p$ total cross-sections and propose a phenomenological
ans\"atz for virtual photon cross-sections. We point out that the good
 agreement with data 
obtained with the Eikonal Minijet Model should not hide the 
many uncertainties buried in the impact parameter distribution.
A model obtained from Soft Gluon Summation is
briefly discussed and its application to hadronic cross-sections is shown.
\end{abstract}
\section{Introduction}
This talk will be a short review of the status of the calculation
of total cross-sections using a  QCD driven mini-jet model\cite{emmus},
with particular emphasis on recent measurements and results of
theoretical calculations of $\gamma \gamma$ cross-sections.

\begin{figure}
\centerline{\epsfig{file=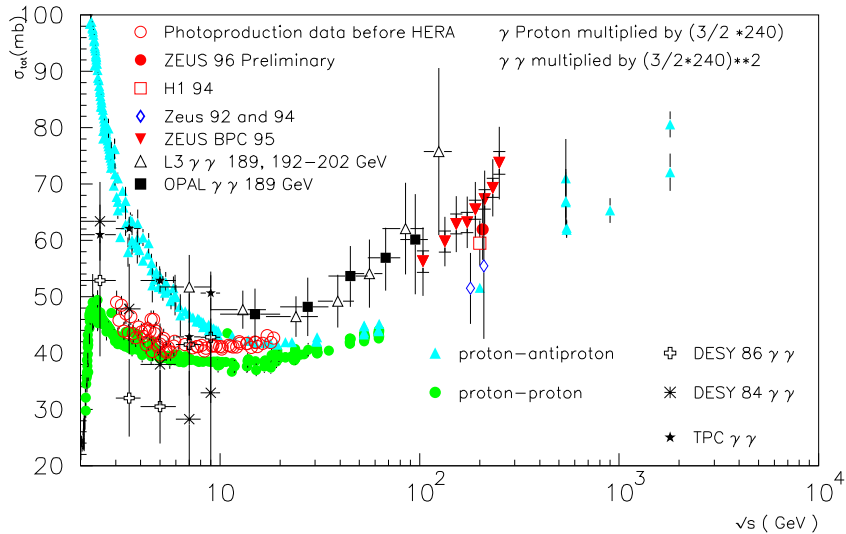,width=12cm}
}
\caption{Energy dependence of $\sigma^{\rm tot}_{ab}$ for various 
choices of $a,b$ as indicated in the figure. The cross-sections for 
the photon-induced processes have been scaled as indicated on the figure.}
\label{allxsec}
\end{figure}

One of the  aim of QCD is  to calculate cross-sections for hadronic 
processes.  We find that the level of available experimental information on
total hadronic cross-sections  has now reached a stage  so as to allow, 
for the first time,   definite progress in the calculation of 
the one quantity which has so far  escaped a complete quantitative 
understanding in a QCD framework, namely the total
hadronic cross-section.  We have a complete set of processes, 
$p p$, $p{\bar p}$, $\gamma p$ and $\gamma \gamma$ measured in a common 
energy range, $\sqrt{s}=1\div 100\ GeV$, with the purely hadronic processes 
measured up to $\sqrt{s}\approx 3\times {10^5} \ GeV$ \cite{martin}. The 
latter allows for very good parametrizations in a large energy range, 
the other two cross-sections to test the QCD content of hadrons versus  
the one in the photons. In Figure \ref{allxsec}
we show a compilation of all presently available photon and proton
total cross-sections, scaling them so as to be able to compare
one to each other. 
For $\gamma \ p$ processes, the scale factor  is
the product of a Quark Parton Model factor $3/2$ multiplied by a Vector Meson 
Dominance type factor $1/240$\cite{emmus}, for $\gamma \gamma$ we just square 
this factor.

The plan of this paper is as follows :
\begin{enumerate}
\item In Sect. II we describe our recent results for $\gamma \gamma$  and 
extrapolate them to $\gamma \ p$
\item In sect. III we discuss a possible ans\"atz on an extension of the
minijet model to $\gamma^* \gamma^*$
\item In Sect. IV we present recent results for $p p$ and $p {\bar p}$,
using  an impact parameter 
distribution for 
partons in the hadrons obtained from the Bloch-Nordsieck summation technique,
and discuss its possible extensions to real and virtual 
photons.
\end{enumerate}

\section{$\gamma \gamma$ and $\gamma\ \lowercase{proton}
$}

\begin{figure}[htb]
\centerline{
\epsfig{file=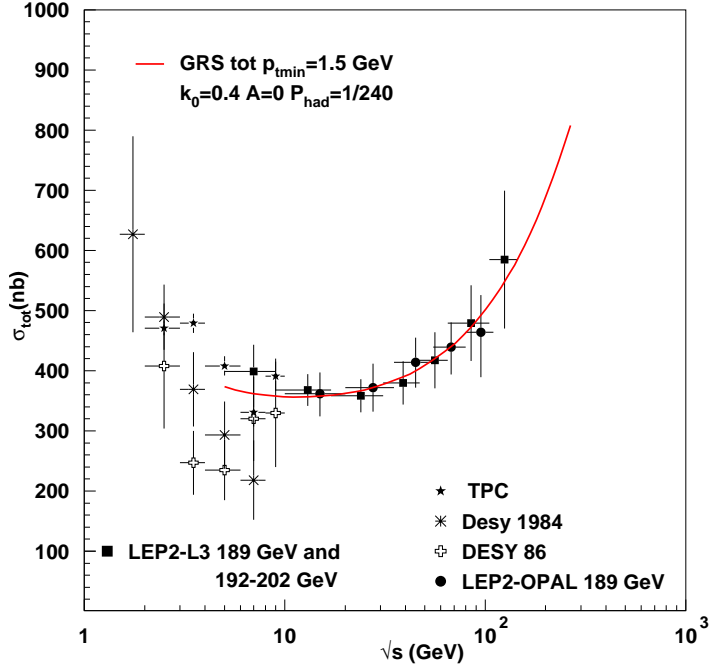,width=10cm}}
\caption{The total 
photon-photon cross-section as described by the EMM (see text)}
\label{maria}
\end{figure}

\begin{figure}[htb]
\centerline{
\epsfig{file=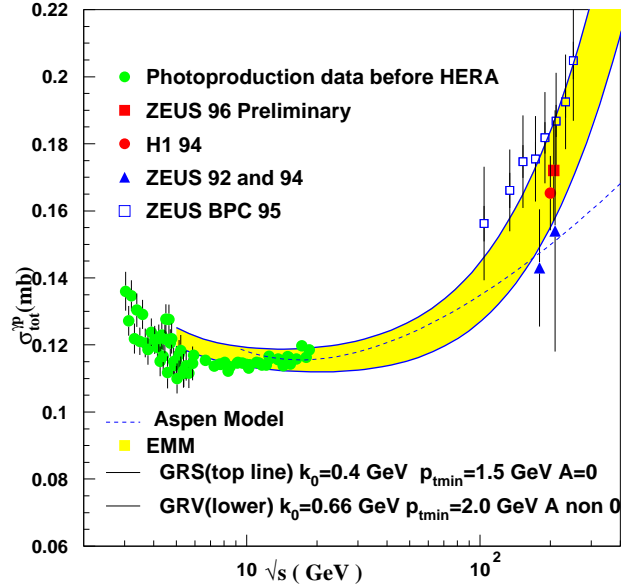,width=10cm}}
\vspace{1cm}
\caption{Comparison between the eikonal minijet model predictions and
data for total $\gamma \ p$ cross-section as well as BPC data extrapolated from
DIS\protect\cite{DIS}. Predictions from \protect\cite{aspen} are also shown.}
\label{camille}
\end{figure}

While for quite some time the photon-photon data at LEP 
exhibited a discrepancy between different collaborations 
(although within the experimental
errors), we now have two sets of data points  which are in 
excellent agreement with
each other. A theorist can then start his/her work.
We show in Figure \ref{maria} the description of L3\cite{L3}
 and OPAL data\cite{OPAL} using the Eikonal Minijet Model (EMM).
The theoretical context in which this curve was obtained is
discussed in \cite{emmus,zeitus} : the EMM uses the eikonal approximation
\cite{eikminijets} to calculate the total cross-section, i.e.
\begin{equation}
\sigma_{tot}=2P_{had}^{ab}\int d^2{\vec b}[1-e^{i\chi(b,s)}]
\label{stotal}
\end{equation}
and approximates the eikonal function neglecting  $\Re e\chi(b,s)$ and
putting
\begin{equation}
2 \Im m\chi(b,s)=n(b,s)=A(b) 
[\sigma_{soft}(s)+{{\sigma_{jet}(s,p_{tmin})}\over{P_{had}^{ab}}}]
\label{nbs}
\end{equation}
where the QCD calculable jet cross-section is the quantity which drives 
the rise \cite{therise} in all total cross-sections. This function is 
defined as
\begin{equation}
\sigma_{jet}=\int d^2{\vec p}_t {{d\sigma^{QCD}}\over{d^2{\vec p}_t}}=
\sum\int dx_1 dx_2 f^{i/a}(x_1)f^{j/b}(x_2) \int 
d^2{\vec p}_t  
{{d{\hat \sigma}^{ij,kl}}\over{d^2{\vec p}_t}}
\end{equation}
where ${\hat \sigma}$ is the parton-parton cross-section for the
subprocess $i j \rightarrow k l$, the sum runs over
     {i,j,k,l,=parton type} and the integration covers a 
region from
a minimum $p_t$ to the entire phase space. The quantity $\sigma_{jet}$
is higly dependent upon the regulator $p_{tmin}$ and one of the aims
of a complete QCD calculation is to eliminate this dependence. Presently this
is not yet possible, but one can nonetheless expect
$p_{tmin}$ to be the smallest momentum exchanged between
partons such that perturbative QCD can be applied, namely not less than 1 GeV 
and probably not more than 2 GeV. The parameter $P_{had}^{ab}\equiv
P_{had}^{a} P_{had}^b$ is
in principle energy dependent and  $P_{had}^a$ can be interpreted as the
probability for particle $a$ to behave like a hadron, i.e. 
$P_{had}^{hadron}\equiv 1$ and,  typically,
$P_{had}^{\gamma}\approx  {\cal O}(\alpha)$ \cite{ladinsky,halzen}.  
Phenomenologically, in
order to obtain a description of the measured total cross-sections,
there are other input quantities which need to be fixed, namely
$\sigma_{soft}$ and the b-distribution function $A(b)$. One
way to proceed has been, previously, to determine all  the
parameters from the process $\gamma p$. However, the EMM
model is not working very well for the proton case, as
discussed in \cite{ff2}, probably because of uncertainties in
the hadronic transverse momentum distributions, and we have opted, in
this note, for a different approach, namely we obtain the
parameters from $\gamma p$, extrapolate them to $\gamma
\gamma$ varying them within at most 
10\% to fit the $\gamma \gamma$ cross-section, and then revert back and 
see
what the best description of $\gamma \gamma$ will produce, when applied to 
photoproduction.
The result of this
{\it modus operandi} is presented here. For Figure \ref{maria},
we have chosen to describe  the partonic matter distribution inside
the hadrons through a function inspired by the pion electromagnetic form factor,
namely
\begin{equation}
A(b)={{1}\over{(2\pi)^2}}
\int d^2 {\vec q} e^{i{\vec q}\cdot {\vec b}}
({{k_0^2}\over{q^2+k_0^2}})^2
\label{aff}
\end{equation}
The scale $k_0$ has been let to vary, according to an
intrinsic transverse momentum ans\"atz \cite{emmus}. Thus,
Figure \ref{maria} corresponds to  $k_0=0.4\ GeV$, 
$P_{had}^\gamma=1/240,\ \sigma_{soft}=(21+{{42}\over{s}})\ mb$, 
$p_{tmin}=1.5\ GeV$ and GRS \cite{GRS}
type densities for the photon. Next we extrapolate this curve to
$\gamma p$ processes, putting
$\sigma_{soft}^{\gamma p}=3/2 \sigma_{soft}$, the proton form factor
with dipole type expression instead of  one of the photon type monopole
expressions, GRV type densities for the proton \cite{GRV} and GRS for the 
photon.
The result is the upper curve shown in Figure \ref{camille} and compared with 
old 
and recent data \cite{HERAZ,HERAH1,DIS,camil}. For
completeness and as a reference to our previous work, we also show
a band, with the lowest curve corresponding to GRV densities for both
proton and photon, $p_{tmin}=2\ GeV$, $k_0=0.66\ GeV$ for the photon,
same $P_{had}^{\gamma}$ as the upper curve, and
$\sigma_{soft}^{\gamma p}=(31+{{10}\over{\sqrt{s}}}+{{38}\over{s}})
\ mb$. 

\section{The case of the virtual photon}
Hadronic interactions of the virtual photon in $e^+e^-$  and $\gamma \gamma$
processes have been a subject of interest for some time now~\cite{virt}. 
The appearance of  experimental data on jet production with virtual photons 
at HERA~\cite{virzeus} and total $\gamma^* \gamma^*$ cross-sections
at LEP~\cite{l3virt}, has given added impetus to develop a model 
which will describe the total cross-sections for virtual photons, 
especially the 'resolved' part~\cite{torbjorn}.  

\begin{figure}[htb]
\centerline{
\epsfig{file=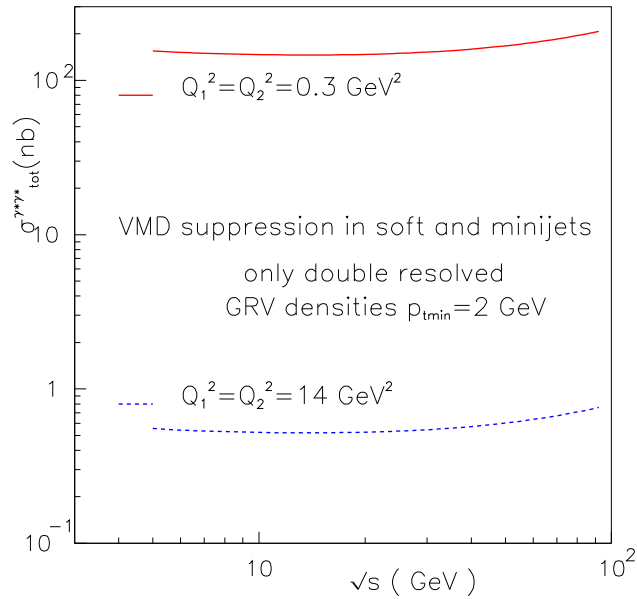,width=10cm}}
\vspace{1cm}
\caption{The EMM predictions 
for virtual photons cross-section. $Q^2_i$ corresponds
to the virtual photon mass}
\label{virtual}
\end{figure}

We want to develop a model to understand these cross-sections in the 
context of EMM. We propose that the virtual photon description is the same 
as for the case of real photons except that the intrinsic transverse 
momentum ans\"atz is complemented with a factor inspired by Extended 
Vector Meson Dominance\cite{bernd}. At high $Q^2_{\gamma}$, a factor like
\begin{equation}
A((b,Q^2_1,Q^2_2)={{m_\rho^2}\over{m_\rho^2+Q^2_1}}
{{m_\rho^2}\over{m_\rho^2 +Q^2_2}} A(b)
\label{EVMD}
\end{equation}
will
suppress the part of the cross-section which
has a hadronic content. In Eq.(\ref{EVMD}),
$P_{had}^\gamma(Q^2_\gamma=0)$ is the usual factor,
 $\approx 1/240$, obtained from Vector meson Dominance
and discussed in many papers\cite{zeitus,ladinsky,halzen}.
In Figure \ref{virtual} we show our predictions for the
hadronic content of the $\gamma^* \gamma^*$ cross-section,
excluding for the time being the direct and single resolved
contribution. Similar results can also be obtained using  
factorization \cite{martinstar} in the context of the Aspen\cite{aspen}
model.

\section{The Bloch Nordsieck Model for the impact parameter distribution}
\begin{figure}[htb]
\vspace{1.5cm}
\centerline{\epsfig{file=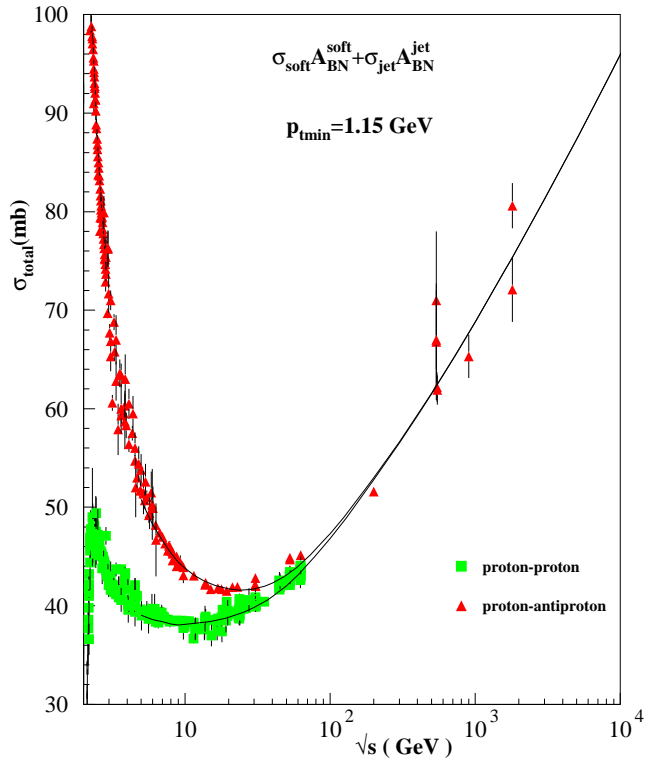,width=10cm}}
\vspace{2cm}
\caption{Total proton-proton and proton-antiproton cross-sections,
compared with an eikonal minijet model which incorporates soft gluon summation}
\label{regge}
\end{figure}

It is evident from the previous discussion, and we have
pointed this out in many papers, that it is not possible to
have a QCD description of the total cross-section without
understanding the transverse momentum distribution
of partons in the colliding hadrons, or, in the eikonal
language, without understanding the
impact parameter distribution. We have attempted
such description, and will show in the following
the main highlights. Some very recent results 
in the case of proton proton and proton antiproton are shown in
Figure \ref{regge},
where the two curves have been obtained using the eikonal formula
of 
Eq.(\ref{stotal}), the jet cross-sections with GRV densities and $p_{tmin}=1.15
\ GeV$, 
$\sigma_{soft}^{pp}=48\ mb$, $\sigma_{soft}^{p {\bar p}}=48 (1+{{2}\over
{\sqrt{s}}}) mb$. Unlike the curves obtainable in a straightforward
application of the EMM with b-distribution determined 
by the proton form factor, this figure uses an s-dependent b-distribution
given by
\begin{equation}
A(b,s)={{\int d^2 {\vec q} e^{i {\vec b} \cdot {\vec q}}\Pi(q,s)}\over
{ (2\pi)^2 \Pi(0,s)}}=
{
{
e^{-h(b,s)}
}\over{
\int d^2{\vec b}
  e^{-h(b,s)}
}}
\end{equation}
where $\Pi(q,s)$ is the transverse momentum distribution of colliding
partons generated by 
soft gluon radiation from the initial state and, to leading order,
\begin{equation}
h(b,s)={{8}\over{3\pi}}\int_0^{q_{max}} \alpha_s(k_t)
 {{dk_t}\over{k_t}}
[1-J_0(b k_t)]
\log{
{
q_{max}+\sqrt{q_{max}^2-k_t^2}
}\over{
q_{max}-\sqrt{q_{max}^2-k_t^2}
}}
\label{hb}
\end{equation}
 The 
functions $h(b,s)$ and $A(b,s)$ are energy dependent inasmuch as the 
kinematics of a given 
subprocess determine how much energy is available to single soft
gluon emission. The energy dependence appears
 through the quantity $q_{max}$ which is a function of 
the c.m. energy $\sqrt{s}$ and $p_{tmin}$. In addition 
$h(b,s)$ depends upon the single
soft gluon distribution function $\alpha_s(k_t) dk_t/k_t$. Since 
$\Pi(q,s)$ is summed over all  possible gluon distributions, the
infrared divergence is of course cancelled, there remains, however, the
problem of evaluating $\alpha_s$ in the infrared limit. We have used
a phenomenological ans\"atz, namely a singular but integrable $
\alpha_s$ and the result  shown in Figure \ref{regge} is discussed
in \cite{ff2} for the rising part of the total proton cross-sections.
 The main characteristics of this treatment is
that, as the minijet cross-section rises with energy,  soft gluon 
emission produces an acollinearity of the partons and reduces the probability 
of collisions. This affects the cross-sections in two ways : at low energy it
produces a very soft decrease in $\sigma^{pp}$ and contributes to the faster
decrease in $\sigma^{p{\bar p}}$,  at high energy it
tames the rise due to $\sigma^{jet}$. It is then possible to
have a very small $p_{tmin}$ to see the very beginning
of the rise around $10\div 20 \ GeV$, without having too large a cross-section
when the energy climbs into the TeV range and beyond.

The above discussion  illustrates also the way to study virtual and real photon
 interactions : the aim is to obtain an energy and momentum dependence in
the impact parameter distribution
through the kinematics characterizing real or virtual photon process,
as it has been done for the proton. At high $Q^2_{\gamma}$, one can expect 
$q_{max}$ in Eq.(\ref{hb}) to increase with $Q^2_{\gamma}$, thus producing the 
same 
suppression effect that high c.m. energy  values have in
the softening of the rise due to the jets. The overall effect
may be similar to the EVMD factor discussed in Sect. III.
Work is in progress to apply the Bloch-Nordsieck formalism to these collisions.

\section{conclusion}
We have discussed total cross-sections for protons and photons, real and 
virtual,
using QCD calculable   mini-jets  cross-sections as the physical process which
 drives the rise of all total cross-sections.  We have used the eikonal 
formalism to unitarize
the cross-section and discussed possible ways to reduce the arbitrariness 
introduced
in this formalism by the impact parameter distribution. A QCD model using
soft gluon summation has been presented and compared with data for purely
proton processes.
\section{acknowledgement}
Two of us, A.G. and G.P., acknowledge the support of EEC TMR-CT98-0169.


\begin{thebibliography}{99}
\bibitem{emmus} A. Corsetti, R.M. Godbole and  G. Pancheri, 
{\it Phys.Lett.} {\bf B435}, 441 (1998).
\bibitem{martin}
M.M. Block, F. Halzen and T. Stanev, {\it Phys.Rev.} {\bf D62},  077501 (2000). 
M.M. Block, F. Halzen, G. Pancheri and T. Stanev,  hep-ph/0003226, {\it 25th 
Pamir-Chacaltaya Collaboration Workshop}, Lodz, Poland, November 1999.
\bibitem{L3} L3 Collaboration,
Paper 519 submitted to {\it ICHEP'98}, Vancouver, July 1998.
M. Acciari et al., {\it Phys. Lett.} {\bf B 408}, 450 (1997); L3 Collaboration,
A. Csilling, {\it Nucl. Phys. Proc. Suppl.} {\bf B82}, 239 (2000).
L3 Note 2548, Submitted to the {\it OSAKA Conference}.
\bibitem{OPAL}OPAL Collaboration. 
G. Abbiendi et al., {\it Eur. Phys. J. } {\bf C14}, 199 (2000). 
F. Waeckerle, {\it Nucl. Phys. Proc. Suppl.} 
{\bf B71}, 381 (1999), {\it Multiparticle Dynamics 1997} Eds. G. Capon, V. Khoze, 
G. Pancheri and A. Sansoni;
Stefan S\"oldner-Rembold, hep-ex/9810011, To appear in  the proceedings
of the {\it ICHEP'98}, Vancouver, July 1998. 
\bibitem{zeitus} R.M. Godbole and  G. Pancheri, e-print Archive 
hep-ph/0010104.
\bibitem{eikminijets}
L. Durand and H. Pi, {\it Phys. Rev. Lett.} {\bf 58}, 58 (1987).
A. Capella, J. Kwiecinsky, J. Tran Thanh, {\it Phys. Rev. Lett.} {\bf 58},  
2015 (1987).
M.M. Block, F. Halzen, B. Margolis, {\it Phys. Rev.} {\bf  D 45}, 839 (1992).
A. Capella and J. Tran Thanh Van, {\it Z. Phys.} {\bf C 23}, 168 (1984).
P. l`Heureux, B. Margolis and P. Valin, {\it Phys. Rev.} {\bf D 32}, 1681 (1985).
\bibitem{therise} 
D. Cline, F. Halzen and J. Luthe, {\it Phys. Rev. Lett.} {\bf 31}, 491 (1973).
G. Pancheri and C. Rubbia, {\it Nucl. Phys.} {\bf A 418}, 117c (1984).
T.Gaisser and F.Halzen, {\it Phys. Rev. Lett.} {\bf 54}, 1754  (1985).
G.Pancheri and Y.N.Srivastava, {\it Phys. Lett.} {\bf B 158}, 402 (1986).
\bibitem{ladinsky}
J.C. Collins and G.A. Ladinsky, {\it Phys. Rev.} {\bf D 43}, 2847 (1991).
\bibitem{halzen}
R.S. Fletcher , T.K. Gaisser and F. Halzen, {\it Phys. Rev.} {\bf D 45}, 377 
(1992); erratum {\it Phys. Rev.} {\bf D 45}, 3279 (1992).
\bibitem{ff2}  A. Grau, G. Pancheri and Y.N. Srivastava,
{\it Phys. Rev.} {\bf D60}, 114020 (1999).
\bibitem{GRS} M. Gl\"uck, E. Reya and I. Schienbein, {\it Phys. Rev.} 
{\bf D60}, 054019 (1999), Erratum-ibid. {\bf D 62}, 019902 (2000).
\bibitem{GRV}
M. Gl\"uck, E. Reya and A. Vogt, {\it Zeit. Physik} {\bf C 67}, 433 (1994).
\bibitem{HERAZ} ZEUS Collaboration, {\it Phys. Lett.} {\bf B 293}, 465 (1992); 
{\it Zeit. Phys.} {\bf C 63}, 391 (1994).
\bibitem{HERAH1} H1 Collaboration, 
{\it Zeit. Phys.} {\bf C 69}, 27 (1995).
\bibitem{DIS}  ZEUS Collaboration, J. Breitweg et al., DESY-00-071, e-print 
Archive: hep-ex/0005018.
\bibitem{camil} ZEUS Collaboration (C. Ginsburg et al.), Proc. 8th 
International Workshop on Deep Inelastic Scattering, April 2000, Liverpool,
to be published in World Scientific.
\bibitem{aspen} M.M. Block, E.M. Gregores, F. Halzen and  G. Pancheri,
 {\it Phys. Rev.} {\bf D58}, 17503 (1998); 
M.M. Block, E.M. Gregores, F. Halzen and G. Pancheri,
{\it Phys. Rev.} {\bf D60}, 54024 (1999).
\bibitem{virt}  See for example, M. Drees and R.M. Godbole,  
{\it Phys. Rev.}, {\bf D 50}, 3124 (1994), e-print Archive: hep-ph/9403229;
{\it Proceedings of PHOTON--95, incarporating the Xth International workshop on
Gamma-Gamma collisions and related processes} , Sheffield,
April 6-10, 1995, pp. 123-130, e-print Archive: hep-ph/9506241.
\bibitem{virzeus}
C. Adloff {\it et al.}, H1 Collaboration, 
{\it Phys. Lett.}, {\bf B 415}, 418 (1997), e-print Archive: hep-ex/9709017;
{\it Eur. Phys. J.}, {\bf C 13}, 397 (2000), e-print Archive: hep-ex/9812024.
\bibitem{l3virt} L3 Collaboration, M. Acciari {\it et al.}, 
{\it Phys. Lett.},{\bf B453}, 333 (1999).
\bibitem{torbjorn}
Ch. Friberg and T. Sj\"ostrand, {\it Eur. Phys. J.}, {\bf C 13}, 151 (2000),
e-print Archive: hep-ph/9907245; {\it JHEP},{\bf 09}, 10 (2000) 
e-print Archive:  hep-ph/0007314.
\bibitem{bernd}
B. Surrow, DESY-THESIS-1998-004;
A. Bornheim, In the  Proceedings of the {\it LISHEP International
School on High Energy Physics}, Brazil, 1998,  hep-ex/9806021.
\bibitem{martinstar} M.M. Block, Invited Talk presented at ISMD2000.
\end{thebibliography}
\end{document}